\documentclass[lettersize,journal]{IEEEtran}
\usepackage{amsmath,amsfonts}
\usepackage{algorithmic}
\usepackage{algorithm}
\usepackage{array}
\usepackage[caption=false,font=normalsize,labelfont=sf,textfont=sf]{subfig}
\usepackage{textcomp}
\usepackage{stfloats}
\usepackage{url}
\usepackage{verbatim}
\usepackage{graphicx}
\usepackage{cite}
\usepackage{color}
\usepackage{isomath} 
\usepackage{colortbl} 
\usepackage{xcolor}
\usepackage{booktabs}
\usepackage{amssymb}
\usepackage{multirow}
\usepackage{makecell}
\usepackage[switch]{lineno}
\usepackage[normalem]{ulem}
\begin{document}
\title{A cross-modal pre-training framework with video data for improving performance and generalization of distributed acoustic sensing}

\author{Junyi~Duan, Jiageng~Chen, and Zuyuan~He,~\IEEEmembership{Senior Member,~IEEE}
\thanks{This work was supported by National Natural Science Foundation of China (NSFC) under Grant 62435004, and 62005163. (\textit{Corresponding author: Zuyuan He.})
}
\thanks{The authors are with the State Key Laboratory of Photonics and Communications, Shanghai Jiao Tong University, Shanghai 200240, China (e-mail: duanjunyi2021@sjtu.edu.cn; zuyuanhe@sjtu.edu.cn).}}

\markboth{}%
{Shell \MakeLowercase{\textit{et al.}}: A Sample Article Using IEEEtran.cls for IEEE Journals}


\maketitle

\begin{abstract}
Fiber-optic distributed acoustic sensing (DAS) has emerged as a critical Internet-of-Things (IoT) sensing technology with broad industrial applications.
However, the two-dimensional spatial-temporal morphology of DAS signals presents analytical challenges where conventional methods prove suboptimal, while being well-suited for deep learning approaches.
Although our previous work, DAS Masked Autoencoder (DAS-MAE), established state-of-the-art performance and generalization without labels, it is not satisfactory in frequency analysis in temporal-dominated DAS data.
Moreover, the limitation of effective training data fails to address the substantial data requirements inherent to Transformer architectures in DAS-MAE.
To overcome these limitations, we present an enhanced framework incorporating short-time Fourier transform (STFT) for explicit temporal-frequency feature extraction and pioneering video-to-DAS cross-modal pre-training to mitigate data constraints. 
This approach learns high-level representations (e.g., event classification) through label-free reconstruction tasks.
Experimental results demonstrate transformative improvements: 0.1\% error rate in few-shot classification (90.9\% relative improvement over DAS-MAE) and 4.7\% recognition error in external damage prevention applications (75.4\% improvement over from-scratch training). 
As the first work to pioneer video-to-DAS cross-modal pre-training, available training resources are expanded by bridging computer vision and distributed sensing areas.
The enhanced performance and generalization facilitate DAS deployment across diverse industrial scenarios while advancing cross-modal representation learning for industrial IoT sensing.
\end{abstract}


\section{Introduction}
\IEEEPARstart{O}ver the last two decades, the Internet of Things (IoT) has pursued the vision of bridging the physical and digital worlds through interconnected networks of devices and systems, which aims to enable pervasive, intelligent automation and data-driven decision-making across diverse applications \cite{rose2015internet,li2015internet}.
Realizing this vision critically depends on deploying highly capable, intelligent sensors that can continuously gather multidimensional physical signals (e.g., acoustic, vibration, strain) while performing advanced processing.
Conventional electromechanical sensors face limitations for large-scale monitoring, including deployment cost, spatial discontinuity, low tolerance to harsh conditions, and power demands.
In contrast, fiber-optic distributed acoustic sensing (DAS) technology emerges as a transformative solution \cite{DAS, DAS2}.
Based on phase-sensitive optical time-domain reflectometry ($\mathrm{\Phi}$-OTDR) \cite{phai-otdr}, a DAS interrogator analyzes the phase of Rayleigh backscattering light along an optical fiber, effectively converting the sensing fiber into tens of thousands of continuous nodes to measure dynamic strain.
This enables distributed sensing with coverage over hundred kilometers, while offering advantages such as high spatial resolution, power-free operation, and adaptability to harsh environments.
These benefits have driven widespread DAS deployment in  IoT applications, including earthquake monitoring \cite{eq1,eq2}, oil and gas pipeline monitoring \cite{pipeline1, pipeline2}, and railway monitoring \cite{railway1, railway2}.  
However, DAS generates spatial-temporal data (often referred to as waterfall plots), presenting unique processing challenges for intelligent interpretation.
Each sensing node outputs a time-series strain signal.
Combining these signals across the fiber forms a two-dimensional (2D) matrix where one axis represents distance (channel) and the other represents time. 
This mechanism creates data fundamentally dominated by temporal dynamics, exhibiting characteristics distinct from natural images and more analogous to acoustic streams. 
Consequently, conventional image processing techniques are often ineffective for learning meaningful representations of underlying physical processes from raw waterfall plots.
New deep learning approaches are therefore essential to learn spatial-temporal representations directly from large volumes of raw DAS data. 

Current deep learning methods for waterfall plot analysis have progressed through several distinct stages of development. 
Initial approaches simplified the problem by reducing 2D waterfall plots to 1D temporal signals, employing conventional architectures including 1D-ResNet \cite{yi2023intelligent}, CNNs \cite{yang2023using,wu2021intelligent}, and LSTM networks \cite{kayan2023intensity}. 
While these methods incorporated sophisticated signal processing techniques, e.g., hidden Markov models \cite{wu2019dynamic,tejedor2019contextual} or short-time Fourier transform (STFT) \cite{pan2022time}, their fundamental limitation lies in discarding the spatial correlations inherent in DAS.
This simplification not only eliminates the possibility of cross-channel validation but also imposes degradations on both algorithm robustness and classification accuracy.
Alternative methods have treated waterfall plots as conventional 2D images \cite{ma2022mi,zhao2021markov}, typically employing standard 2D convolutional architectures like U-Net \cite{van2021self}. 
However, these approaches fail to account for the structural differences between natural images (dominated by spatial relationships) and waterfall plots (characterized by temporal evolution patterns). 
Consequently, these methods often produce suboptimal spatial-temporal representations that struggle to capture the essential dynamics of the underlying phenomena.
More advanced hybrid approaches have attempted to address these limitations through sequential processing pipelines that first extract 1D temporal features before performing spatial fusion \cite{li2024exploiting,wu2020novel,li2020fiber}. 
While these methods demonstrate improved performance through explicit temporal-spatial modeling, the sequential processing paradigm disrupts symmetry between temporal and spatial information, resulting in a compromised representation quality.
Our previous work, DAS masked autoencoder (hereafter referred to as DAS-MAEv1) \cite{duan2025mae}, addressed these limitations by implementing the Transformer architecture \cite{2017attention} within a self-supervised masked autoencoder \cite{MAE}.
This approach enables simultaneous processing of spatial-temporal information in waterfall plots while leveraging unlabeled data, significantly outperforming semi-supervised baselines with superior generalization capabilities.
However, two critical challenges remain: First, the embedding layers in Transformers are inadequate to capture short-time frequency characteristics and lose critical spectral information. 
Second, the Transformer's requirements of massive data necessitate expansion of training data sources to realize its representation learning potential.

On the other hand, recent advances in speech processing have demonstrated remarkable success in applying Transformer architectures to spectrograms to learn temporal-frequency representation \cite{wang2023neural,radford2023robust}.
From this perspective, we propose enhanced DAS masked autoencoder (hereafter refer to as DAS-MAEv2) for waterfall plot representation learning that effectively addresses above two key challenges. 
This approach begins by transforming each spatial channel's time-series signal into a 2D spectrogram using STFT, which converts an original 2D waterfall plot into a 3D spatial-temporal-frequency tensor through spatial stacking of these spectrograms. 
Since the structural similarity between our 3D DAS tensor and 3D video data, we incorporate video pre-training to simultaneously address the Transformer's data requirements while exploring domain adaptation possibilities. 
The DAS-MAEv2 architecture implements an asymmetric autoencoder design, where both encoder and decoder components are Video Vision Transformers (ViViT) \cite{vivit}, adapted for waterfall plot analysis.
During pre-training, the model is optimized through a self-supervised reconstruction task, recovering 3D waterfall plots from inputs with 90\% randomly masked portions. 
This pre-training is done in two stages: first, the pre-training is implemented on video data to establish fundamental spatial-temporal representations, and followed by domain-specific adaptation using waterfall plots. 
Remarkably, the pre-trained model demonstrates strong clustering capabilities, automatically grouping representations from the same event types (e.g., walking, digging) without any label supervision.
Experimental validation on an open benchmark dataset \cite{dataset} demonstrates the framework's effectiveness. 
In few-shot learning scenarios, DAS-MAEv2 achieves a 0.1\% error rate (90.9\% relative improvement over DAS-MAEv1) and reaches only a 5.8\% error rate with only 15 labeled samples per class. 
Furthermore, the pre-trained model consistently outperforms from-scratch training across varying dataset sizes with over 73.9\% relative improvements in practical external damage prevention applications.
These results demonstrate  model's high performance and strong transferability to novel applications and event types.

\section{Methods}
\label{sec2: methods}
\subsection{Distributed acoustic sensing and waterfall plots}
\label{subsec: DAS}
\begin{figure*}[th]
\begin{center}
\begin{tabular}{c}
\includegraphics[width=0.96\linewidth]{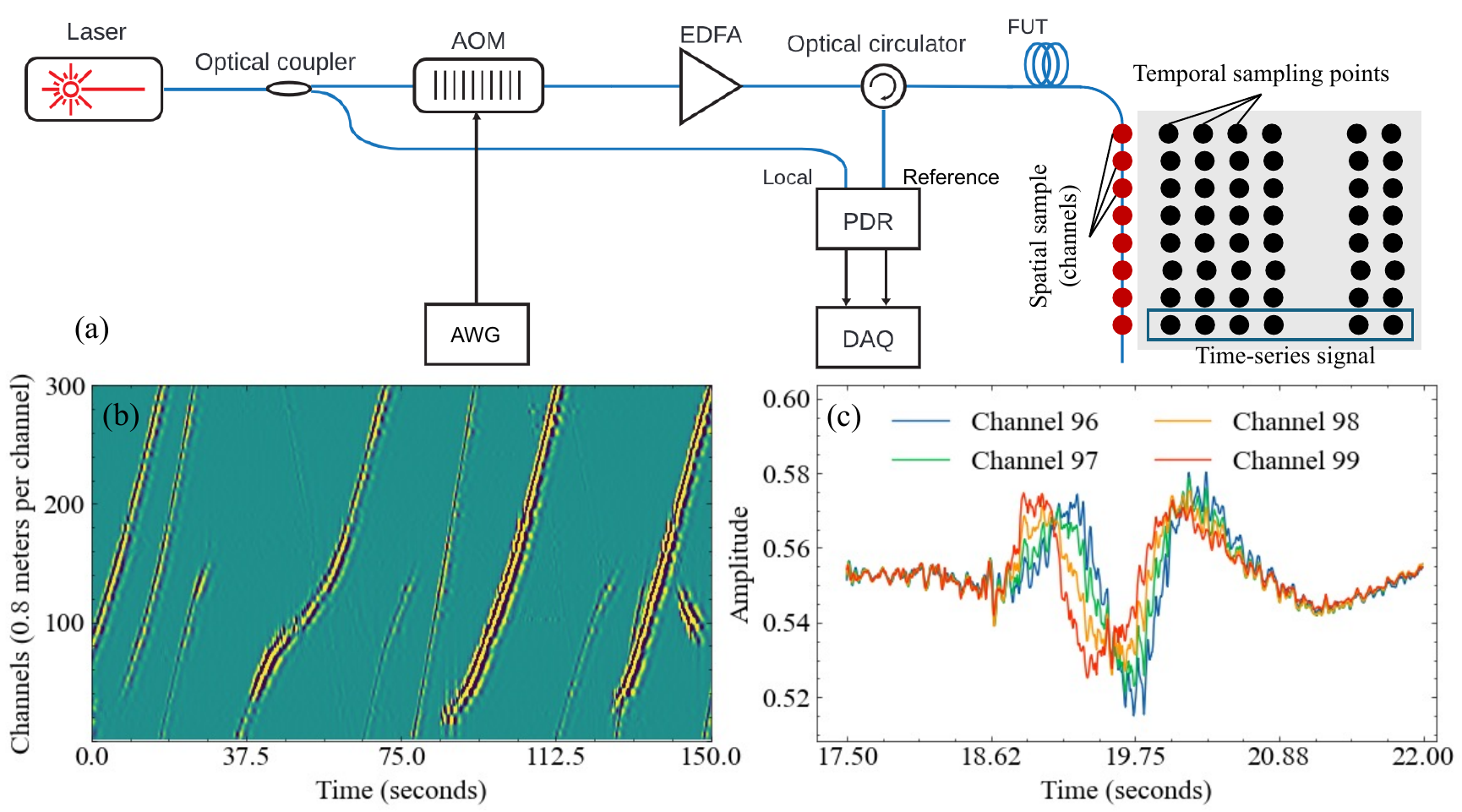}
\end{tabular}
\end{center}
\caption 
{ \label{fig1}
Distributed acoustic sensing system and its acquired waterfall plot. 
(a) Scheme of DAS.
The system is composed of a laser, an acousto-optic modulator (AOM), an arbitrary waveform generator (AWG), an erbium-doped fiber amplifier (EDFA), a polarization diversity receiver (PDR), a data acquisition card (DAQ), an optical coupler, an optical circulator, and optical fibers (blue lines). 
A spatial segment of fiber under test (FUT) is a spatial sample or channel, which is equal to an acoustic sensor, such as a microphone.
The time-series signals from each spatial sample are stacked into a 2D matrix, forming a waterfall plot.
(b) A waterfall plot records the vehicle motion trajectories along a roadside.
Each channel represents 0.8 meters.
(c) The time-series signals at adjacent channels from Fig. \ref{fig1}(b).
} 
\end{figure*}

Fig. \ref{fig1}(a) illustrates the working principle of a DAS system.
A coherent continuous-wave (CW) laser source emits light, which is split into two branches via an optical coupler.
In the interrogation branch (upper branch), an acousto-optic modulator (AOM), driven by modulation signals from an arbitrary waveform generator (AWG), generates linear frequency-modulated (LFM) laser pulses. 
These pulses are amplified by an erbium-doped fiber amplifier (EDFA) and then directed through an optical circulator into the sensing fiber or the fiber under test (FUT). 
Rayleigh backscattering (RBS) light from the FUT propagates back through the circulator to the signal port of a polarization diversity receiver (PDR) \cite{ren2015theoretical}.
The local branch (lower path) provides the local input for coherent detection at the PDR's local port.
A data acquisition card (DAQ) samples the heterodyne signals from the PDR, followed by digital signal processing (DSP) on a computer.
The external strain $\epsilon$, applied to the FUT, varies the fiber refractive index.
Consequently, it introduces the RBS phase difference as:
\begin{equation}
\label{eq1}
    \mathrm{\Delta} \mathrm{\Phi}_L(\epsilon)=2\beta(n+C_\epsilon)\epsilon L=\frac{\epsilon L}{K_\mathrm{\Phi}},
\end{equation}
where $\beta=2 \pi/\lambda$ is the wave vector in vacuum, $n$ is the fiber core refractive index, $C_\epsilon$ is the photo-elastic coefficient, $L$ is the spatial differential distance, and $K_\mathrm{\Phi}$ is sensitivity coefficient.
For a laser wavelength of 1550 $nm$ and a fiber core refractive index $n=$ 1.46, the sensitivity coefficient $K_\mathrm{\Phi}$ is around 110.37~$\ n\epsilon \cdot m/rad$ \cite{chen2019108,dong2016quantitative}.
Equation \ref{eq1} establishes a linear relationship between the strain (vibration waveform) and the RBS phase difference. 
Crucially, the phase difference remains zero outside vibration-active regions, enabling precise vibration localization.
The hundred-kilometer-long FUT is transformed into tens of thousands of sensing nodes (spatial samples) by DAS.
Each sensing node measures the strain and outputs a corresponding time-series signal.
By combining the time-series signals from multiple sensing nodes, a 2D spatial-temporal matrix is formed as a waterfall plot.

Fig. \ref{fig1}(b) and (c) illustrate the vehicle motion trajectories recorded by DAS along a roadside, manifesting as non-stationary spatial-temporal signals across adjacent channels and sampling points.
Unlike 2D images, the horizontal axis of waterfall plots represents time, while the vertical axis corresponds to spatial position along the sensing fiber (FUT). 
Moreover, the temporal sample rate is significantly higher than the spatial sample rate in DAS measurements.
For instance, the temporal sample rate is 200 samples per second, compared to the spatial sample rate of 1.25 samples per meter (i.e., 0.8 meters per channel) in Fig. \ref{fig1}. 
This results in an asymmetric spatial-temporal information density, where temporal information dominates in information entropy.
Furthermore, the signal amplitude in DAS measurements depends on both vibration intensity and fiber-medium coupling efficiency, where the coupling efficiency varies significantly along the sensing fiber in practice.
As shown in Fig. \ref{fig1}(b) (signals in 37.5 to 75.0 second interval), channels before 200 exhibit stronger fiber-ground coupling than subsequent channels, causing amplitude variations while preserving spectral pattern similarity.
This physical constraint necessitates prioritized analysis of frequency-domain features when processing time-series signals in waterfall plots.
Fig. \ref{fig1}(c) displays the time-series signals from channels 96 to 99. 
These signals contain rich waveform and frequency information, and adjacent channels exhibit substantial redundancy (similar waveforms or waveform coherence \cite{seismic}).
Prior studies \cite{wu1dcnn, wu2021pattern} demonstrated that focusing on temporal feature extraction within waterfall plots substantially improves learned representations. 
However, exclusive reliance on temporal information, while neglecting spatial correlations, could yield suboptimal results.
This distinctive characteristic of waterfall plots establishes a unique spatial-temporal coupling, fundamentally different from that observed in natural images.

\subsection{DAS-MAEv2}
\label{subsec: DASMAE}

\begin{figure*}[htbp]
\begin{center}
\begin{tabular}{c}
\includegraphics[width=0.96\linewidth]{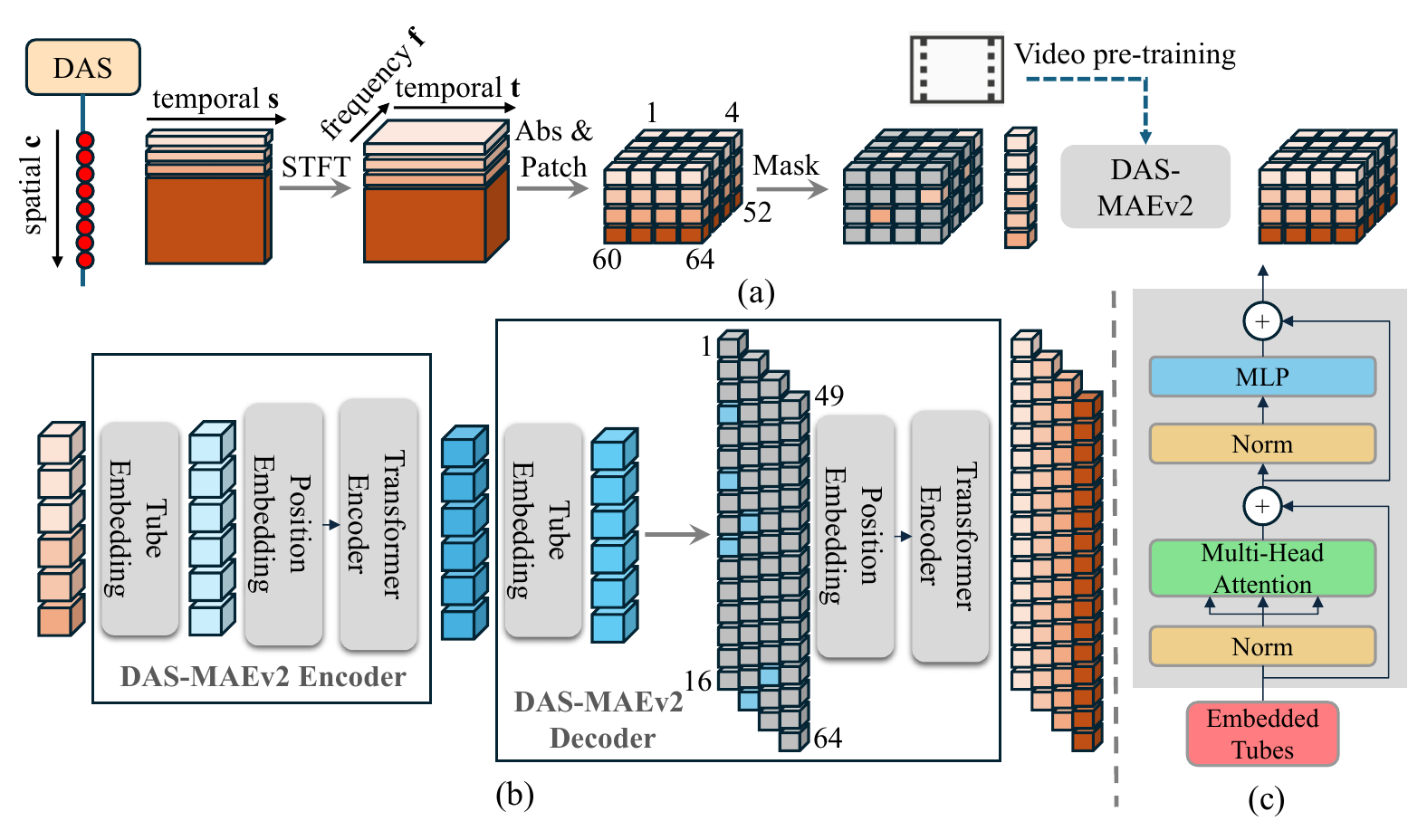}
\end{tabular}
\end{center}
\caption 
{ \label{fig: DASMAE}
Architecture of DAS-MAEv2 in pre-training. 
(a) The framework of DAS-MAEv2. Its hyperparameters are given in Table \ref{tab:DAS-MAE}.
(b) The inside modules of DAS-MAEv2.  
It employs the Video Vision Transformers (ViViT) \cite{vivit} for both the encoder and decoder, each including a tube embedding layer, a position embedding layer, and a Transformer encoder. 
It's important to note that the encoder is deliberately designed with a larger size compared to the decoder (see Table \ref{tab:DAS-MAE}). 
(c) The structure of a Transformer encoder (depth of $L$, head of $H$) \cite{vit}. It consists of $L$ Transformer blocks, each including multi-head attention ($H$ heads), a MLP block, and norm layers.
} 
\end{figure*} 

Fig. \ref{fig: DASMAE}(a) illustrates the DAS-MAEv2 framework for self-supervised masked autoencoding of waterfall plots.
A DAS waterfall plot is represented as $\mathbf{x} \in \mathbb{R}^{C \times S}$, where $C$ denotes the number of spatial channels and $S$ represents temporal sampling points.
For each channel $i$ of $\mathbf{x}$, we compute the absolute value of the STFT transformation with a window length of $L$, hop length of $L$, and FFT size of $L$, padded with $M$ zeros per window. 
This yields the spatial-temporal-frequency data $\mathbf{X} \in \mathbb{R}^{C \times T \times F}$, where $T=\left\lfloor S/L \right\rfloor$ is the number of time frames, and $\lfloor \cdot \rfloor$ is the floor function.
$F=L/2$ is the number of frequency bins, satisfying the relationship $\left\lfloor S/T \right\rfloor = F-M$.
The input $\mathbf{X}$ is partitioned into $N$ non-overlapping spatial-temporal-frequency tubes $\{\mathbf{X}_i \in \mathbb{R}^{C_p \times T_p \times F_p}\}_{i=1}^N$, where $ N = \left\lfloor C/C_p \right\rfloor \times \left\lfloor T/T_p \right\rfloor \times \left\lfloor F/F_p \right\rfloor$.
$C_p$, $T_p$, and $F_p$ denote the number of channels, time frames, and frequency bins in a tube, respectively.
During pre-training, we implement aggressive random masking by removing 90\% of tubes ($N_m=\lfloor 0.9N \rfloor$) following a uniform distribution (referred to as "random sampling").
The remaining $N_v=N - N_m$ visible tubes $\mathbf{X_{{m}^{c}}}$ (where $\mathbf{{m}^{c}}$ denotes the complement of the mask set $\mathbf{{m}}$) are encoded by the DAS-MAEv2 encoder into latent representations.
A lightweight decoder estimates the masked tubes $\mathbf{\hat{X}_m}$ from representations and mask tokens.
The reconstruction objective minimizes the normalized mean square error (MSE) across masked regions within the entire dataset $\mathcal{X}$:
\begin{equation}
L_{rec}= \mathbb{E}_{\mathbf{X} \sim \mathcal{X}} \left[\frac{1}{N_m}\sum_{i=1}^{N_m}\left \| \mathbf{X}_\mathbf{m}^{(i)}- {\mathbf{\hat{X}_m}}^{(i)} \right \|^2 _{2}\right].
\label{eq:training loss}
\end{equation}
The 90\% mask ratio (the ratio of removed tubes) largely eliminates the waterfall plot's redundancy and prevents interpolation solutions.
The lightweight decoder design forces the encoder to generate high-level representations from data points.

As illustrated in Fig. \ref{fig: DASMAE}(b), the DAS-MAEv2 framework employs a Video Vision Transformer (ViViT) \cite{vivit} for both its encoder and decoder. 
The encoder begins by processing visible input tubes $\mathbf{X_{{m}^{c}}} \in \mathbb{R}^{N_v \times C_p \times T_p \times F_p \times D_i}$ (where $D_i=1$ is an additional data dimension for network processing) through a 3D convolutional layer that projects the tube dimension $D_i$ to $D_e$.
Then the tubes are flattened as $\mathbb{R}^{N_v \times C_p \times T_p \times F_p \times D_e} \rightarrow \mathbb{R}^{N_v \times (C_p \cdot T_p \cdot F_p) \times D_e}$.
To preserve positional information, standard 1D learnable positional encodings \cite{2017attention} are incorporated before the sequence of embedded tubes undergoes processing through $L_e$ stacked Transformer blocks.
Each Transformer block, demonstrated in Fig. \ref{fig: DASMAE}(c), comprises multi-head attention with $H_e$ heads, MLP layers, and normalization operations.
The encoder ultimately yields high-level representations $\mathbf{Z_{{m}^{c}}} \in \mathbb{R}^{N_v \times (C_p \cdot T_p \cdot F_p) \times D_e}$.
The decoder operates on these representations $\mathbf{Z_{{m}^{c}}}$ by first applying tube embedding (a linear projection) to dimension $D_d$, then reconstructing the complete set of $N$ tokens through insertion of a shared learnable mask token $\mathbb{R}^{1 \times (C_p \cdot T_p \cdot F_p) \times D_d}$ at masked positions.
The 1D position embeddings are also added to maintain spatial-temporal-frequency relationships.
The sequence is then transformed through $L_d$ Transformer blocks ($H_d$ heads) before final reshaping to match the original input dimensions $\mathbb{R}^{N \times C_p \times T_p \times F_p \times D_i}$.
Complete architectural specifications, including tensor dimensions and hyperparameter values are systematically presented in Table \ref{tab:DAS-MAE}.

\begin{table*}[htbp]
\caption{Pre-training framework of DAS-MAEv2}
\begin{center}      
\resizebox{\textwidth}{!}{
\begin{tabular}{cccccc}
\Xhline{1pt}
\textbf{Module/Variable} & \textbf{Sub-module} & \textbf{Layer} & \textbf{Hyperparameter} & \textbf{Output size/Size} & \textbf{Value}\\
\hline
$\mathbf{x}$  & - & - & - & $\mathbb{R}^{C \times S \times D_i}$ & $C_p=2$ \\ \cline{1-5} 
$\mathbf{X}$ (STFT \& patched) & - & - & - & $\mathbb{R}^{N \times C_p \times T_p \times F_p \times D_i}$ & $T_p=16$ \\ \cline{1-5} 
$\mathbf{X_{{m}^{c}}}$ (masked) & - & - & -  & $\mathbb{R}^{N_v \times C_p \times T_p \times F_p \times D_i}$ & {$ F_p=16$}  \\
\cline{1-5} 
\multirow{3}{*}{Encoder} &  Tube embedding & 3D Conv & $\mathbf{kernel}=\mathbf{stride}=C_p \times T_p \times F_p$ & $\mathbb{R}^{N_v \times (C_p \cdot T_p \cdot F_p) \times D_e}$ & $N=216$ \\
& Position embedding & - & - & $\mathbb{R}^{N_v\times (C_p \cdot T_p \cdot F_p) \times D_e}$ & $N_v=21$  \\
& Transformer encoder & Transformer block & $\mathbf{depth}=L_e, \mathbf{head}=H_e$ & $\mathbb{R}^{N_v \times (C_p \cdot T_p \cdot F_p) \times D_e}$ & $D_e=384$ \\
\cline{1-5} 
$\mathbf{Z_{{m}^{c}}}$ & Representations (masked) & - & - & $\mathbb{R}^{N_v \times (C_p \cdot T_p \cdot F_p) \times D_e}  $ & $D_d=192$
\\
\cline{1-5} 
\multirow{3}{*}{Decoder} & Tube embedding & Linear & $\mathbf{in\_channel}=D_e, \mathbf{out\_channel}=D_d$ & $\mathbb{R}^{N_v \times (C_p \cdot T_p \cdot F_p) \times D_d}$ &  $L_e=12$ \\
& Position embedding & - & - & $\mathbb{R}^{N \times (C_p \cdot T_p \cdot F_p) \times D_d}$ &  $H_e=6$ \\
& Transformer encoder & Transformer block & $\mathbf{depth}=L_d, \mathbf{head}=H_d$ & $\mathbb{R}^{N \times (C_p \cdot T_p \cdot F_p) \times D_d}$ & $L_d=4$\\ 
\cline{1-5} 
$\mathbf{\hat{X}}$ & - & - & - & $\mathbb{R}^{N \times C_p \times T_p \times F_p \times D_i}  $ & $H_d=3$
\\\hline
\Xhline{1.1pt}
\end{tabular}
}
  \label{tab:DAS-MAE}
\end{center}
\end{table*}

DAS-MAEv2 significantly advances beyond DAS-MAEv1 by incorporating STFT to enable joint temporal-frequency analysis.
This approach addresses critical limitations in the original DAS-MAEv1 framework, where the raw temporal series are directly time-embedded via employing a DAS-ViT architecture (analogous to Vision Transformers for images). 
This time-embedded overemphasized temporal amplitude variations (theoretically linear to vibration intensity in ideal DAS conditions) while being limited by spatial variations in fiber-medium coupling efficiency.
These variations introduce non-linear distortions between the measured amplitude and the actual vibration energy. 
The STFT transformation converts 2D spatial-temporal matrices into interpretable 3D spatial-temporal-frequency tensors that explicitly reveal frequency-domain patterns while maintaining temporal fidelity.
This transformation provides more robust features, particularly through instantaneous frequency shifts that are less susceptible to coupling artifacts.
Consequently, the usage of STFT transformation facilitates more effective representation learning for DAS.
The STFT results, i.e., 3D waterfall plots (spatial-temporal-frequency), require replacing DAS-ViT with ViViT for 3D tensor processing.
The structural similarity between 3D waterfall plots (spatial-temporal-frequency) and video data (temporal-spatial-spatial) inspires a dual-stage pre-training approach, which involves initial masked reconstruction on video data and is followed by further pre-training on 3D waterfall plots.
Comprehensive ablation studies in Section \ref{sec4:abl} demonstrate the individual contributions of STFT integration, dual-stage pre-training, and mask-reconstruction hyperparameter settings to learned representation quality. 

\subsection{Dual-stage pre-training strategy}
The DAS-MAEv2 framework ($\sim$23M parameters) employs a dual-stage pre-training approach. 
In the first stage, the model is pre-trained on video data, where we directly use the trained parameters from prior work \cite{videomae} since it is not the primary focus of our current study. 
The second stage involves domain-specific adaptation through continued pre-training on an open-access waterfall plot dataset contributed by X. Cao et al. \cite{dataset}. 
This comprehensive dataset 
consists of approximately 15000 samples, with each sample containing temporal sequences of 12 channels $\times$ 10000 sampling points, acquired at sample rates of either 12.5 kHz or 8.0 kHz. 
The dataset encompasses six distinct event categories: background noise, digging, knocking, watering, shaking, and walking.
The training set contains about 12000 samples (approximately 2000 samples per event category), while the testing set comprises 3000 samples (roughly 500 samples per event category).

For the domain adaptation stage, we utilize only the raw data without any class labels during pre-training. The training configuration employs a batch size of 64 across 500 epochs, using the AdamW optimizer \cite{adamw} with a cosine decay learning rate schedule (initial learning rate of $10^{-3}$, 40-epoch warmup period) \cite{cos_lr}. 
The complete pre-training process requires approximately 5 hours on an NVIDIA RTX 4090 GPU. 
The pre-trained model is able to achieve real-time inference (4 ms @ RTX 4090), which is suitable for field deployment.

\subsection{Evaluating the pre-trained DAS-MAEv2}
\label{subsec: eva}

To assess the learned representation quality, we evaluate DAS-MAEv2's performance on a classification task with labeled dataset $\mathcal{D}=\{(\mathbf{x}^{(k)}, \mathbf{y}^{(k)}) \}_{k=1}^K$ where $\mathbf{y} \in \{0, 1\}^M$ is a one-hot encoded label over $M$ event classes.
Given an input $\mathbf{x} \in \mathbb{R}^{C \times S}$, the pre-trained encoder $\mathbold{E}$ generates latent representations $\mathbf{Z}\in \mathbb{R}^{N \times (C_p \cdot T_p \cdot F_p) \times D_e}$ \textit{without masking}.
Consequently, the representation length changes from $N_v$ to $N$.
These representations are aggregated via averaging along the first dimension:
\begin{equation}
\bar{\mathbf{Z}}=\underset{\text{(1st dim)}}{\mathrm{Mean}}\big(\mathbold{E}[\mathrm{STFT}(\mathbf{x})]\big),
\end{equation}
where $\mathrm{STFT}(\cdot)$ is the short-time Fourier transform. 
The average representation $\bar{\mathbf{Z}}$ is then mapped to class probabilities $\mathbf{\hat{y}}$ via two protocols.

Linear probing uses a lightweight linear classifier $\mathbold{D}$, which is a single linear layer, to project average representation $\bar{\mathbf{Z}}$ to class probabilities, while the encoder $\mathbold{E}$ remains frozen:
\begin{equation}
\arg \min _{\mathbold{D}} \mathbold{L}_{cls}\left[\mathbf{y}, \mathbold{D} \left(\underset{\text{(1st dim)}}{\mathrm{Mean}}\big(\mathbold{E}[\mathrm{STFT}(\mathbf{x})]\big)\right)\right],
\end{equation}
where $\mathbold{L}_{cls}(\cdot)$ is the cross-entropy loss \cite{krizhevsky2012imagenet} for classification.
This method evaluates whether class-relevant representations emerge as linearly separable clusters in the latent space, which directly reflects the encoder’s reconstruction-driven representation quality.

Fine-tuning attempts to use task-specific nonlinear features and optimizes both the encoder $\mathbold{E}$ and classifier $\mathbold{D}$ (a single linear layer):
\begin{equation}
\arg \min _{\mathbold{E},\mathbold{D}} \mathbold{L}_{cls}\left[\mathbf{y}, \mathbold{D}\left(\underset{\text{(1st dim)}}{\mathrm{Mean}}\big(\mathbold{E}[\mathrm{STFT}(\mathbf{x})]\big)\right)\right].
\end{equation}
This method demonstrates the model’s adaptability. 
It leverages minimal task-specific training (far fewer epochs than supervised training from scratch) to maximize transfer performance.

\section{Experiements}
\label{sec: pre-train}

\subsection{Visualization of learned representations}
\begin{figure*}[htbp]
\begin{center}
\begin{tabular}{c}
\includegraphics[width=0.9\linewidth]{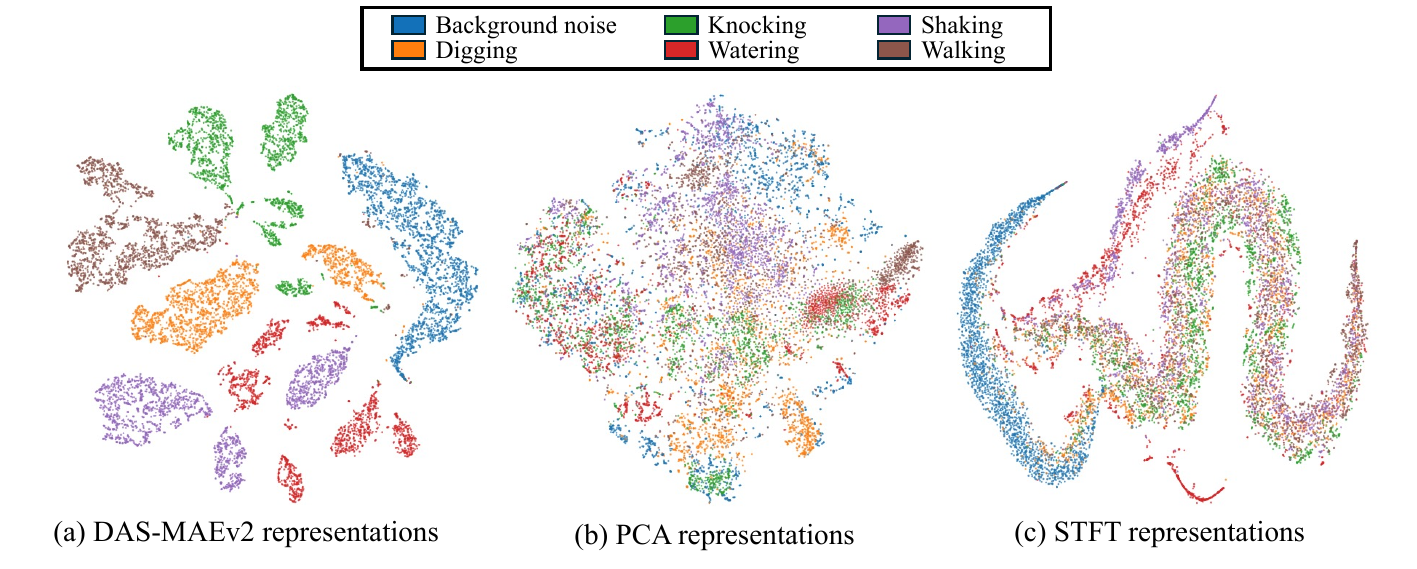}
\end{tabular}
\end{center}
\caption 
{\label{fig: visual}
The t-SNE visualization of representations learned by DAS-MAEv2, principal component analysis (PCA), and STFT transformation on the open dataset \cite{dataset}.
Labels are used to assign colors and evaluate the quality of learned representations.
(a) Averaged representations $\bar{\mathbf{Z}}$ from DAS-MAEv2.
(b) Representations learned by PCA.
(c) Representations obtained from the STFT transformation of raw data.
} 
\end{figure*}

We first conducted a qualitative evaluation of DAS-MAEv2's learned representations through visualization using t-distributed stochastic neighbor embedding (t-SNE) \cite{t-SNE}. 
As a nonlinear dimension reduction technique, t-SNE effectively preserves clustering patterns from high-dimensional spaces in low-dimensional projections.
For our analysis, the pre-trained encoder extracted average representations $\bar{\mathbf{Z}}$ from the training dataset, which we further projected to the 2D space using t-SNE with perplexity of 40, learning rate of 2000, and 500 iterations. 
Class labels were used solely for color assignment in Fig. \ref{fig: visual} to enhance visual discrimination.

Fig. \ref{fig: visual}(a) demonstrates that DAS-MAEv2 generates well-separated clusters for all six event categories, with minimal inter-class overlap. 
The background noise class (blue) forms a compact cluster, which is consistent with its stationary nature.
Although dynamic events (walking, knocking, etc.) show intra-class dispersion, our ablation studies in Section \ref{sec4:abl} confirm this can be effectively resolved through supervised fine-tuning and achieve less than 0.1\% error rate.
For comparative evaluation, we contrast these representations with those from principal component analysis (PCA) \cite{PCA}. 
Fig. \ref{fig: visual}(b) reveals PCA's fundamental limitation: its linear projections fail to discriminate between event types, producing completely overlapping clusters. 
While Fig. \ref{fig: visual}(c) shows that STFT-transformed waterfall plots provide some separable spectral representations between background noise and dynamic events (unlike PCA), these representations still cannot achieve the clear inter-class separation demonstrated by DAS-MAEv2 in Fig. \ref{fig: visual}(a). 
These visual comparisons provide empirical evidence that DAS-MAEv2 learns semantically meaningful representations without label supervision.

\subsection{Quality of learned representations}
\label{sec: eva}
To evaluate the learned representations, we compared DAS-MAEv2 against DAS-MAEv1 and the semi-supervised ACNN-SA-BiLSTM (ACAB) model \cite{MTssl} on the open dataset \cite{dataset}.
Since the ACAB model was originally trained under the Mean Teacher framework \cite{MT} with limited labeled data, we ensured an equitable evaluation by fine-tuning all DAS-MAE versions under identical few-shot learning constraints (i.e., using the same labeled data subsets). 
For fairness, all models were fine-tuned using AdamW (initial learning rate=10$^{-5}$, weight decay=0.05, 4 warm-up epochs) with a cosine scheduler for 50 epochs \cite{adamw}.
We quantify improvements via the Relative Improvement (RI):
\begin{equation}
    \mathrm{RI}(\mathrm{A}, \mathrm{B})=\frac{\mathrm{ER}_\mathrm{A}-\mathrm{ER}_\mathrm{B}}{\mathrm{ER}_\mathrm{A}} \times 100\%.
\end{equation}
where $\mathrm{ER}_\mathrm{A}$ and $\mathrm{ER}_\mathrm{B}$ denote the error rates of models A and B, respectively.
Here, $\mathrm{RI}(\mathrm{A},\mathrm{B})$ explicitly measures the percentage reduction in error rate when replacing model A with model B. 
Note that RI is asymmetric: $\mathrm{RI}(\mathrm{A}, \mathrm{B})\neq \mathrm{RI}(\mathrm{B}, \mathrm{A})$, as it normalizes improvement relative to the baseline model (A in this case).

\begin{table*}[htbp]
\caption{Few-shot learning error rate of different models}
\label{tab:ERR}
\begin{center}
\begin{tabular}{w{c}{1.7cm}w{c}{1.7cm}w{c}{2 cm}w{c}{2cm}w{c}{2cm}w{c}{1.5cm}w{c}{1.5cm}}
\Xhline{1.1pt}
\textbf{Data No.} &  \multicolumn{3}{c}{\textbf{DAS-MAE}} & \multirow{2}{*}{\textbf{ACAB}\cite{MTssl}} & \multicolumn{2}{c}{\textbf{RI}}\\ 
\cline{2-4}
\cline{6-7}
\textbf{(per class)}&  \textbf{v1} \cite{duan2025mae} & \textbf{v2-w/o-video} & \textbf{v2}&  & \textbf{(v1, v2)} & \textbf{(ACAB, v2)}\\
\hline
$15$ & 10.0\% & 7.1\% & \cellcolor{gray!30}5.8\% & 16.5\% & 42.0\% & 64.8\%
\\
$40$ &  3.6\% & \cellcolor{gray!30}1.1\% & \cellcolor{gray!30}1.1\% & 6.2\% & 69.4\% & 82.3\%
\\
$80$ & 2.8\% & 0.6\% & \cellcolor{gray!30}0.3\% & 4.4\% & 89.3\% & 93.2\%
\\

$120$ & 1.3\% & 0.4\% & \cellcolor{gray!30}0.2\% & 3.8\% & 84.6\% & 94.7\%
\\
$205$ & 1.1\% & 0.2\% & \cellcolor{gray!30}0.1\% & 3.1\% & 90.9\% & 96.8\%
\\
\hline
\Xhline{1.1pt}
\end{tabular}
\end{center}
  \label{tab: compare with semi-supervised method}
\end{table*}

The experimental results presented in Table \ref{tab:ERR} demonstrate the superior classification performance of our DAS-MAEv2 model compared to both the previous DAS-MAEv1 version and the semi-supervised ACAB model across various few-shot learning scenarios.
With only 15 labeled samples per class (90 total), DAS-MAEv2 achieves an ER of 5.8\%, representing a significant 64.8\% RI over ACAB's 16.5\% ER and a 42.0\% improvement over DAS-MAEv1's 10.0\% ER. 
This performance advantage becomes even more pronounced as the number of labeled samples increases, where DAS-MAEv2 reaches an exceptional 0.1\% ER with 205 samples per class, corresponding to a 96.8\% RI over ACAB's 3.1\% ER and a 90.9\% RI over DAS-MAEv1's 1.1\% ER. 
The 'v2-w/o-video' variant, which incorporates STFT transformations but excludes video data pre-training, already shows substantial improvements over v1 (e.g., 7.1\% vs 10.0\% ER at 15 samples).
While the full DAS-MAEv2 model with video data further achieves additional performance gains (7.1\% vs 5.8\% ER at 15 samples). 
These results not only validate the effectiveness of our architectural improvements but also highlight DAS-MAEv2's remarkable capability to achieve practical deployment requirements (sub-10\% ER) with minimal labeled data, a threshold that ACAB fails to meet under the same low-data conditions. 
The progressive increase in RI values with more training data (from 42.0\% to 90.9\% for v1 to v2) further demonstrates DAS-MAEv2's superior representation learning ability compared to both its predecessor and the semi-supervised ACAB model.

\subsection{Field evaluation of representation generalization}
\begin{figure*}[h]
\begin{center}
\begin{tabular}{c}
\includegraphics[width=0.85\linewidth]{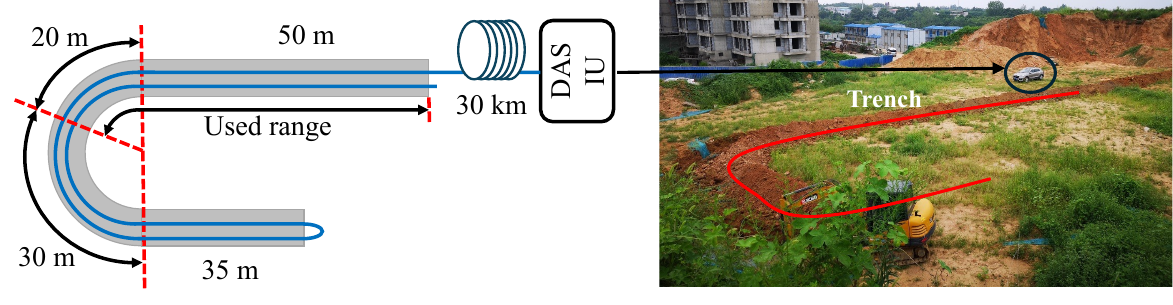}
\end{tabular}
\end{center}
\caption 
{ \label{fig: real scenery}
Experimental layout of the external damage prevention.
The experimental configuration features a U-shaped trench containing the buried sensing fiber.
The approximately 135-meter-long trench maintains an average depth of 1 meter, consisting of three distinct segments: a 50-meter linear section, a 50-meter curved section, and a 35-meter linear section. 
Here, DFVS-850 DAS system was employed with a sample rate of 2 kSa/s and 10-meter spatial resolution. 
The waterfall plot is only captured through the sensing fiber within the specified range, which includes the first 50 meters and the 20 meters of the curved section ('used range').
} 
\end{figure*}

To thoroughly evaluate the generalization capability of learned representations under real-world conditions, we conducted field experiments within an external damage prevention application deployed in Zhengzhou, China. 
The experimental configuration (Fig. \ref{fig: real scenery}) utilized a U-shaped fiber trench containing three geometrically distinct sections.
It was monitored by a DFVS-850 DAS system operating at a sample rate of 2 kSa/s with 10-meter spatial resolution.
We established a carefully designed vibration dataset consisting of eight distinct categories:

(1) \textbf{Background noise (Class 0)}: Baseline recordings captured under static conditions with no active vibration sources, comprising 42 samples.

(2) \textbf{Road roller in motion (Class 1)}:  Pure movement of the road roller along the fiber optic cable path without compaction operations, including 100 samples.

(3) \textbf{Road roller in compaction (Class 2)}: Combined movement and dynamic vertical compaction forces applied by the road roller, with 300 collected samples.

(4) \textbf{Excavator in excavation (Class 3)}: 
Full excavation activities were performed using two different machinery configurations, including both light-duty (6-ton) and heavy-duty (22-ton) excavators. 
This class contains 332 samples.

(5) \textbf{Excavator in motion (Class 4)}: 
Pure movement of excavators along the fiber path, captured in 334 samples.

(6) \textbf{Electric drill in operation (Class 5)}: 
High-frequency vibrational signals generated by handheld electric drilling equipment, represented by 38 samples.

(7) \textbf{Fully-loaded forklift in motion (Class 6)}: 
Movement of a 9-ton forklift operating at maximum cargo capacity (total weight 11 tons), comprising 54 samples.

(8) \textbf{Unloaded forklift in motion (Class 7)}: Movement of the same 9-ton forklift in unloaded configuration, with 186 collected samples.

The complete dataset contains approximately 1400 vibration events in total. 
Notably, this data distribution exhibits significant and intentional class imbalance that reflects real-world operational conditions. 
Excavator-related activities (Classes 3 and 4) occur nearly 10 times more frequently than electric drill operations (Class 5), while background noise recordings (Class 0) are particularly sparse at just 42 samples.
This carefully constructed imbalance serves two critical evaluation purposes: First, it rigorously tests the model's ability to maintain detection sensitivity for rare but operationally important events (such as electric drill operations). 
Second, it challenges the learned representations to resist classification bias toward dominant categories (like excavator activities) that could otherwise lead to overfitting. 
While this distribution disparity presents significant learning challenges, it provides an authentic assessment of the model's capacity to achieve robust features, which is a crucial requirement for practical field deployment.

\begin{table*}[htbp]
\caption{Error rate of DAS-MAEv2 in practical external damage prevention application}
\begin{center}      
\begin{tabular}{w{c}{2.5cm}w{c}{1.7cm}w{c}{1.7 cm}w{c}{2.2cm}w{c}{2cm}w{c}{1.7cm}}
\Xhline{1.1pt}
\textbf{Training data}  & \multicolumn{2}{c}{\textbf{DAS-MAE}} & \textbf{Scratch model} & \multicolumn{2}{c}{\textbf{RI}}\\
\cline{2-3}
\cline{5-6}
\textbf{(portion$|$volume)} & \textbf{v1}\cite{duan2025mae} & \textbf{v2} & \textbf{(v2 structure)} & \textbf{(Scratch, v2)} & \textbf{(v1, v2)} \\
\hline
100\%$|$1100  & 5.0\% & \cellcolor{gray!30} 4.7\% & 19.1\% & 75.4\% & 6.0\% \\
62.5\%$|$700  & 12.4\% & \cellcolor{gray!30} 5.3\% & 28.2\% & 81.2\% & 57.3\% \\
25\%$|$280  & 16.2\% & \cellcolor{gray!30} 9.1\% & 34.8\% & 73.9\% & 43.8\%
\\\hline
\Xhline{1.1pt}
\end{tabular}
  \label{tab: real ps}
\end{center}
\end{table*}

\begin{figure*}[h]
\begin{center}
\begin{tabular}{c}
\includegraphics[width=0.95\linewidth]{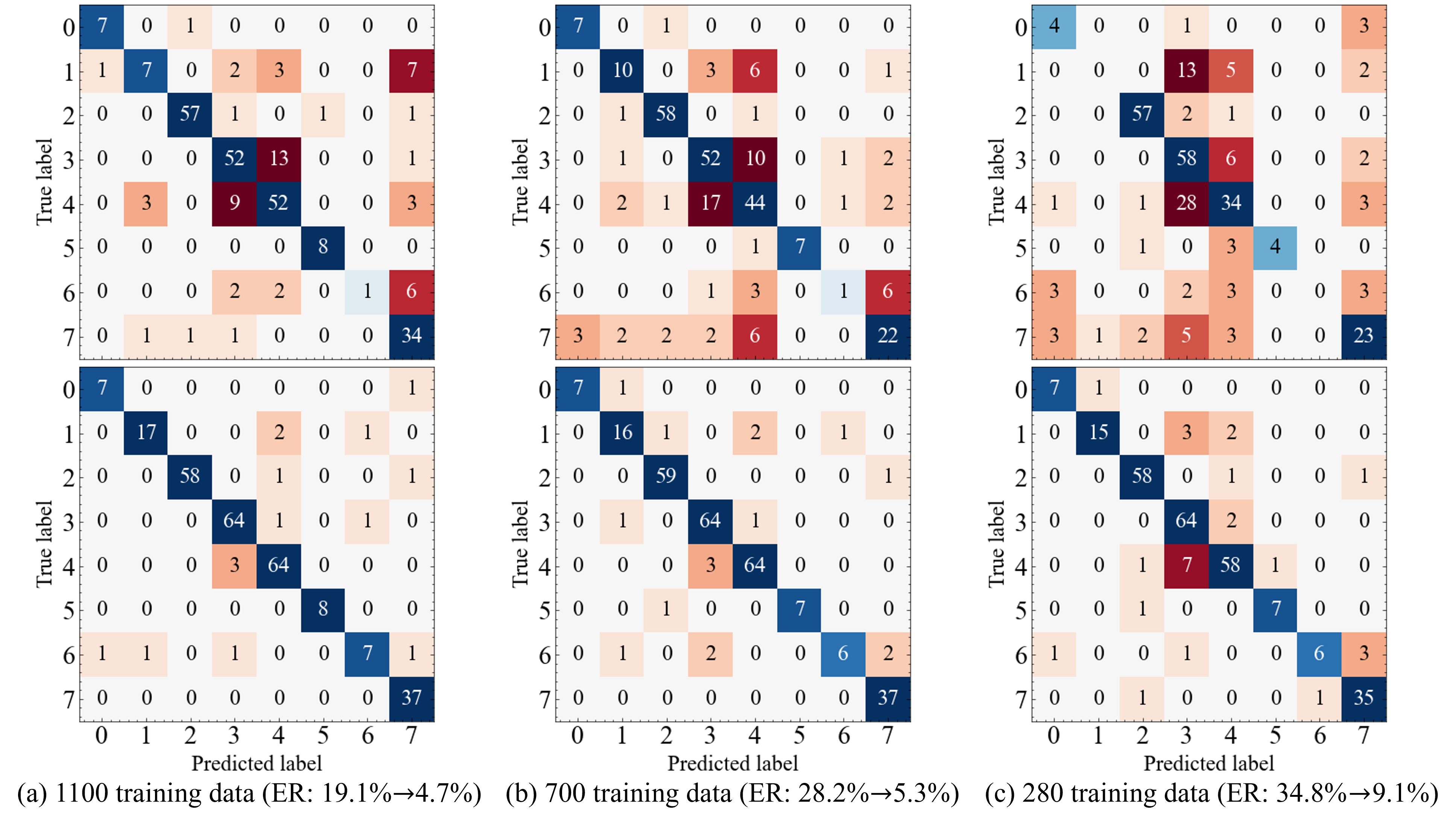}
\end{tabular}
\end{center}
\caption 
{ \label{fig: real cm}
Confusion matrix comparison between the pre-trained model (i.e., DAS-MAEv2) and the scratch model across different training set sizes.
The upper row in each comparison displays results from the scratch model (same architecture but trained from scratch), while the bottom row shows corresponding results from DAS-MAEv2. 
(a) 1,100 training samples: DAS-MAEv2 reduces the error rate (ER) from 19.1\% (scratch model) to 4.7\%.
(b) 700 training samples: DAS-MAEv2 reduces ER from 28.2\% to 5.3\%.
(c) 280 training samples: DAS-MAEv2 reduces ER from 34.8\% to 9.1\%.
} 
\end{figure*} 

The waterfall plot dataset was partitioned into training and testing sets using a 4:1 ratio per event class. 
We evaluated the pre-trained DAS-MAE through fine-tuning on three subsets of the training data: full capacity (100\%, $\sim$1,100 samples), medium capacity (62.5\%, $\sim$700 samples), and low capacity (25\%, $\sim$280 samples), with each configuration trained for 150 epochs.
The results are given in Table \ref{tab: real ps}.
When utilizing the complete training set, the pre-trained model achieved a 4.7\% test ER, surpassing practical deployment requirements by 5.3\%.
This represents a 75.4\% relative improvement (14.4\% absolute improvement) compared to scratch-model training (trained until convergence) and a 6\% relative improvement over DAS-MAEv1. 
As evidenced in Fig. \ref{fig: real cm}(a), the pre-trained DAS-MAEv2 exhibits superior performance across all classes.
It commits significantly more corrections (7 vs. 1) in full-load forklift operation classification (Class 6), demonstrating effective mitigation of class imbalance.
The pre-trained DAS-MAEv2 resolves the confusion between excavator states (excavation vs. motion) that plagues the scratch model.
Under data-limited conditions, the pre-trained model maintained robust performance with error rates of 5.3\% (62.5\% training data) and 9.1\% (25\% training data), consistently outperforming scratch models by $>$70\% relative improvement ($\sim$20\% absolute improvement) and DAS-MAEv1 by $>$40\% relative improvement. 
Fig. \ref{fig: real cm}(b) and (c) illustrate the specific confusion matrices.
These results demonstrate the substantial effectiveness of DAS-MAEv2's pre-training phase and prove the high transferability of our model in different applications with novel event types.

\section{Ablations study}
\label{sec4:abl}
To optimize DAS-MAEv2's configuration for waterfall plot analysis and representation learning, we performed hyperparameter ablation studies.
Using a controlled experimental protocol, we first pre-trained models with individual hyperparameter variations under identical training schedules (Section \ref{sec: pre-train}), then assessed representation quality through standardized linear probing and fine-tuning on the benchmark dataset \cite{dataset}, reporting test-set classification error rate as the quantitative metric. 
Throughout these experiments, we strictly modified only one hyperparameter at a time while keeping all other architectural components fixed to ensure isolated variable analysis.

\begin{table*}[htbp]
\centering
\caption{Ablation studies on DAS-MAEv2  (measured by error rate)}
\label{tab: ab}
\resizebox{\textwidth}{!}{
\begin{tabular}{cc}
    \textbf{(a) Ablation on STFT transformation and video pre-training} & \textbf{(b) Ablation on mask ratio} \\
    \begin{tabular}{cccc}
        \Xhline{1.1pt}
        \multirow{2}{*}{\textbf{DAS-MAE}} & \multirow{2}{*}{\textbf{Mask ratio}} & \textbf{ER of} & \textbf{ER of} \\
         &  & \textbf{linear probing} & \textbf{fine-tuning} \\
        \hline
        v1 from Ref. \cite{ duan2025mae} & 50\% & 2.45\% & 0.32\%\\
        v1-same-params & 50\% & 7.86\% & 2.77\%\\
        v2-w/o-video & 90\% & 0.42\% & 0.16\%\\
        v2 & 90\% & \cellcolor{gray!30} $\mathbf{0.23\%}$& \cellcolor{gray!30} $\mathbf{0.06\%}$\\
        \Xhline{1.1pt}
    \end{tabular}
    \label{tab: ab patch shape}
    &
    \begin{tabular}{ccc}
        \Xhline{1.1pt}
        \textbf{Mask ratio} & \textbf{ER of linear probing} & \textbf{ER of fine-tuning}\\
        \hline
        70\% & 0.35\%  & 0.13\% \\
        80\% &  0.48\% & \cellcolor{gray!30} $\mathbf{0.06\%}$  \\
        90\% &\cellcolor{gray!30} $\mathbf{0.23\%}$ & \cellcolor{gray!30} $\mathbf{0.06\%}$ \\
        95\% & 1.10\% & 0.19\% \\
        98\% & 11.33\% & 0.51\%  \\
        \Xhline{1.1pt}
    \end{tabular} \\
    \textbf{(c) Ablation on mask strategy} & \textbf{(d) Ablation on input data} \\
    \begin{tabular}{ccc}
        \Xhline{1.1pt}
        \textbf{Mask strategy} & \textbf{ER of linear probing} & \textbf{ER of fine-tuning} \\
        \hline
        Random sampling &\cellcolor{gray!30} $\mathbf{0.23\%}$& \cellcolor{gray!30} $\mathbf{0.06\%}$ \\
        Spatial sampling & 4.48\% & 1.01\% \\
        Temporal sampling & 4.09\% & 0.48\% \\
        Frequency sampling & 4.03\% & 0.41\% \\
        \Xhline{1.1pt}
    \end{tabular}
    &
    \begin{tabular}{ccc}
        \Xhline{1.1pt}
        \textbf{Data} & \textbf{ER of linear probing} & \textbf{ER of fine-tuning} \\
        \hline
        Magnitude &\cellcolor{gray!30} $\mathbf{0.23\%}$& \cellcolor{gray!30} $\mathbf{0.06\%}$ \\
        Magnitude \& Phase & 0.81\% & 0.16\% \\
        Real \& Imaginary & 76.01\% & 81.22\% \\
        \Xhline{1.1pt}
    \end{tabular}
\end{tabular}
}
\label{tab: ablation study}
\end{table*}

\subsection{STFT transformation and video pre-training}
\begin{figure*}[h]
\begin{center}
\begin{tabular}{c}
\includegraphics[width=0.55\linewidth]{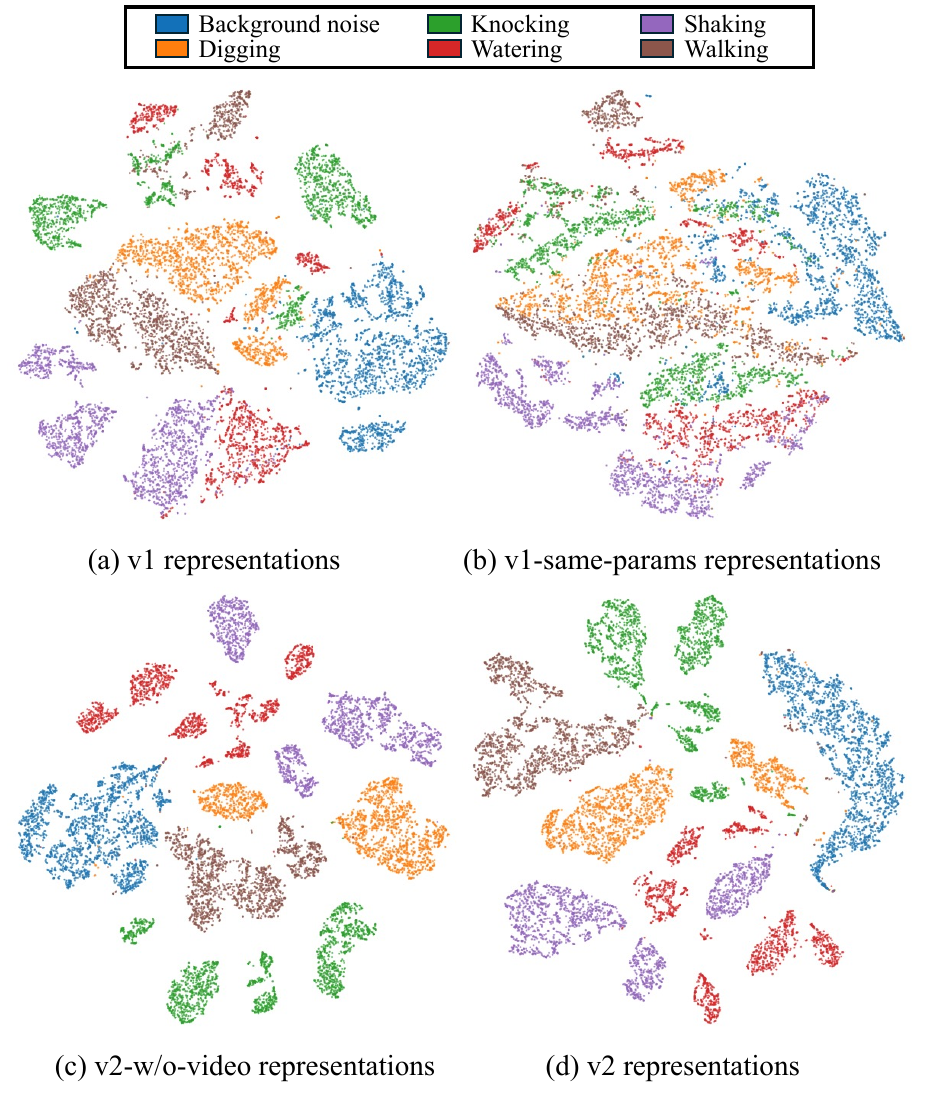}
\end{tabular}
\end{center}
\caption 
{\label{fig: abl_visual}
The t-SNE visualization of representations learned by DAS-MAEv1, v1-same-params (a modified version of v1 using v2's hyperparameter settings), v2-w/o-video (v2 architecture without video pre-training), and v2 on the open dataset \cite{dataset}.
Labels are used for color assignment to evaluate the quality of learned representations.
(a) Representations from DAS-MAEv1.
(b) Representations learned by v1-same-params.
(c) Representations obtained from v2-w/o-video.
(d) Representations from DAS-MAEv2.
} 
\end{figure*}

To thoroughly investigate the impact of STFT transformation and video pre-training on model performance, we conducted a detailed comprehensive comparison across distinct model configurations: (1) the original DAS-MAEv1 \cite{duan2025mae}, (2) a modified version of v1 using v2's hyperparameter settings on Transformer blocks for the same model size as v2 (denoted as v1-same-params), (3) v2 architecture without video pre-training (v2-w/o-video), and (4) the complete DAS-MAEv2 with all enhancements.
The t-SNE visualizations of learned representations from these models on the open dataset are presented in Fig. \ref{fig: abl_visual}.
The visualization results reveal important insights about the models' representation learning capabilities.
In Fig. \ref{fig: abl_visual}(a), the original DAS-MAEv1 shows well-clustered representations where samples from the same event class aggregate together with clear separation between different classes. 
By contrast, the v1-same-params variant in Fig. \ref{fig: abl_visual}(b) demonstrates degraded clustering performance, with noticeable overlaps between distinct event classes, particularly between digging (orange) and walking (brown).
This performance degradation was expected and can be explained by the fundamental architectural differences between v1 and v2. 
The v1 architecture was specifically optimized for capturing spatial-temporal coherence in waterfall plots, while v2's parameter configuration was designed to better handle joint spatial-temporal-frequency coherence.
The benefits of STFT transformation become evident in Fig. \ref{fig: abl_visual}(c), where the v2-w/o-video model shows obvious improvement over both v1 variants. 
The representations exhibit clearer class boundaries than DAS-MAEv1, with distinct separation emerging between digging (orange) and walking (brown) events, as well as between shaking (purple) and watering (red) events. 
This demonstrates the advantage of incorporating frequency-domain information through STFT transformation.
When examining the complete DAS-MAEv2 in Fig. \ref{fig: abl_visual}(d), we observe that the additional video pre-training provides extra benefits. 
The most noticeable improvement appears in the background noise class, whose representations become more tightly clustered. 
For other event classes, the enhancements from video pre-training are less observable. 
Collectively, these visualization results demonstrate that STFT transformation and video pre-training contribute to learning more discriminative representations.

Quantitative results are presented in Table \ref{tab: ab patch shape}(a).
Our experimental results show that the original DAS-MAEv1 architecture achieved error rates of 2.45\% for linear probing and 0.32\% for fine-tuning when using its optimal 50\% mask ratio. 
However, when using the v1-same-params model, we observed increased ER of 7.86\% for linear probing and 2.77\% for fine-tuning. 
This result aligns with the observation in the t-SNE visualization in Fig. \ref{fig: abl_visual}(a) and (b).
The implementation of STFT transformation in the v2-w/o-video model brought substantial performance improvements, reducing the error rates to 0.42\% for linear probing and 0.16\% for fine-tuning. 
These results represent significant relative improvements of 94.7\% and 94.2\%, respectively, compared to the v1-same-params configuration. 
During our experiments, we discovered that the optimal mask ratio for the v2 architecture increased to 90\%, which suggests that the spatial-temporal-frequency 'waterfall plots' exhibit much higher information redundancy compared to raw spatial-temporal waterfall plots. 
We believe this enhanced redundancy stems from STFT's unique ability to explicitly reveal the critical frequency-domain features hidden in the raw data.
When we further incorporated video pre-training to create the full DAS-MAEv2 model, we achieved our best results with error rates of 0.23\% for linear probing and 0.06\% for fine-tuning. 
These figures correspond to additional relative improvements of 45.2\% and 62.5\% over the v2-w/o-video configuration, clearly demonstrating the complementary benefits of both architectural innovations.
The STFT transformation provides superior frequency-aware representation learning capabilities, while the video pre-training contributes valuable transferable feature extraction abilities that further enhance model performance.

\subsection{Mask ratio}
The mask ratio constitutes a critical hyperparameter in DAS-MAEv2 that regulates the difficulty of the reconstruction task and determines the quality of learned representations.
This parameter requires careful balancing, where the task must be sufficiently challenging to prevent the model from relying on simple interpolation strategies, yet not so difficult as to hinder the learning of fundamental signal characteristics. 
This delicate balancing ensures the encoder acquires precisely the necessary information to learn high-quality representations that capture meaningful semantics (e.g., event categories).
As demonstrated in Table \ref{tab: ab patch shape}(b), the optimal representations emerge at an aggressive 90\% mask ratio, yielding minimal error rates of 0.23\% (linear probing) and 0.06\% (fine-tuning). 
Both more conservative (less than 70\%) and excessively aggressive (over 95\%) masking ratios result in suboptimal representations. 
The model exhibits different sensitivity to mask ratios depending on the evaluation approach. 
Fine-tuning demonstrates robust performance across a wide range of mask ratios (70\%-95\%) in this classification task.
However, linear probing shows huge error rate reductions of 0.8\% between 90\% and 95\% mask ratio, and 10\% between 95\% and 98\% ratio. 
This contrast underscores the practical advantage of fine-tuning for achieving stable performance in practical applications.

The 90\% optimal mask ratio for 3D spatial-temporal-frequency 'waterfall plots' presents a notable contrast to the 50\% optimal ratio observed for 2D spatial-temporal waterfall plots. 
This difference indicates that our model can effectively leverage more heavily masked inputs while actually improving performance. 
Previous studies prove that the optimal mask ratio correlates strongly with the inherent redundancy of the input signal. 
For instance, video data (with higher redundancy than images) shows optimal mask ratios of 90\% \cite{videomae} compared to 75\% for images \cite{MAE}. 
Similarly, while 2D waterfall plots achieve best performance at 50\% masking, 3D waterfall plots peak at 90\%, suggesting significantly higher information redundancy in the latter case.
Crucially, since the STFT transformation itself does not introduce extra information, this finding implies that STFT successfully exposes latent information within 2D waterfall plots, making the signal appear more redundant to the DAS-MAEv2 model. 
This phenomenon provides evidence for the effectiveness of incorporating STFT transformation in our framework, as it enables the model to work with less visible inputs while simultaneously improving representation quality.

\subsection{Mask sampling strategy}
\begin{figure*}[t]
\begin{center}
\begin{tabular}{c}
\includegraphics[width=0.8\linewidth]{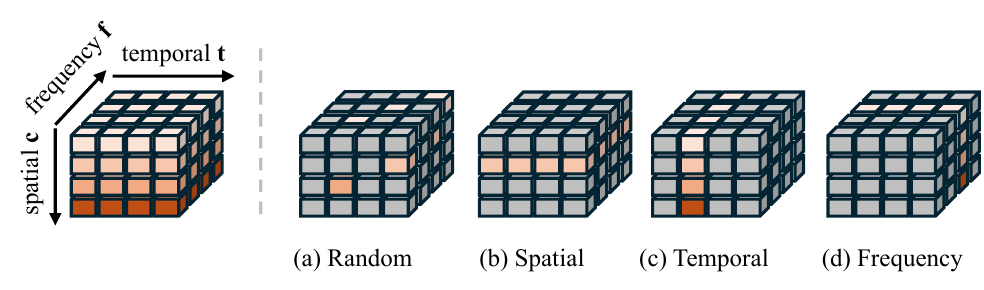}
\end{tabular}
\end{center}
\caption 
{ \label{fig: mask samping}
Mask sampling strategies determine the difficulty of the pre-training reconstruction task, influencing the quality of learned representations. 
(a) Random sampling (our default) masks the tubes following a uniform distribution. 
(b) Temporal sampling removes the entire temporal-wise tubes. 
(c) Spatial sampling removes the entire spatial-wise tubes.
(d) Frequency sampling removes the entire frequency-wise tubes.
} 
\end{figure*}

Table \ref{tab: ab patch shape}(c) systematically evaluates the impact of four distinct mask sampling strategies on DAS-MAEv2's representation learning performance, with visual illustrations of each strategy provided in Fig. \ref{fig: mask samping}. 
The spatial sampling strategy (Fig. \ref{fig: mask samping}(b)) removes entire spatial-wise tubes and compels the model to perform reconstruction using spatial-adjacent temporal-frequency information (i.e., time-series similarity in the waterfall plot). 
This design facilitates learning inter-channel relationships across 2D temporal-frequency data, but it presents challenges for capturing intrinsic temporal-frequency distributions within individual channels.
Under this limitation, DAS-MAEv2 only achieved error rates of 4.48\% (linear probing) and 1.01\% (fine-tuning) with spatial sampling, highlighting the importance of spatial correlations in waterfall plot analysis.
These results underscore the advantages of distributed sensors over traditional single-point sensors.
The spatial correlations enable cross-validation between adjacent sensors, provide complementary signal features unavailable in isolated measurements, and allow for error correction through majority consensus among neighboring channels.

The temporal sampling strategy (Fig. \ref{fig: mask samping}(c)), which eliminates temporal redundancy by masking temporal-wise tubes, demonstrated improved performance with a 4.09\% linear probing error rate (representing a 9\% relative improvement over spatial sampling) and 0.48\% fine-tuning error rate (52\% relative improvement).
Frequency sampling (Fig. \ref{fig: mask samping}(d)), which removes entire frequency-wise tubes, achieved further gains with a 4.03\% linear probing error rate (1\% relative improvement over temporal sampling) and 0.41\% fine-tuning error rate (15\% relative improvement). 
These progressive improvements confirm the dominant role of frequency information in 3D waterfall plot representation learning, where accurate understanding of frequency distributions enables higher-quality feature extraction.

The random sampling strategy in Fig. \ref{fig: mask samping}(a) proved as the optimal approach, achieving superior performance with a 0.23\% linear probing error rate (94\% relative improvement over frequency sampling) and 0.06\% fine-tuning error rate (85\% relative improvement).
The effectiveness of random sampling stems from its ability to simultaneously leverage spatial, temporal, and frequency correlations while introducing beneficial stochasticity that enhances model robustness and prevents overfitting. 
This comprehensive sampling approach allows the model to learn balanced representations that capture all relevant dimensions of the 3D waterfall plot data structure.

\subsection{STFT output data format}
The STFT transformation converts the original 2D waterfall plot data into 3D complex-valued data.
To accommodate our model's requirement for real-valued input, we evaluate three different real-valued formats of STFT results: magnitude-only (absolute values), concatenated magnitude and phase components, and concatenated real and imaginary parts. 
The concatenation operation is performed along the additional input dimension $D_i$.
We conducted a systematic ablation study to evaluate how these formats affect DAS-MAEv2's representation learning capability.
The comparative results for different input representations are presented in Table \ref{tab: ab patch shape}(d).
When using concatenated real and imaginary parts as input, the model achieved worst performance with linear probing and fine-tuning error rates of 76.01\% and 81.22\%, respectively. 
The result indicates the model failed to learn meaningful representations from this input format.
In contrast, using concatenated magnitude and phase components yielded significantly better results, with error rates of 0.81\% (linear probing) and 0.16\% (fine-tuning).
This improvement aligns with established knowledge in time-series signal processing (e.g., speech recognition), where spectral magnitude typically contains more discriminative information than phase components.
Since waterfall plots essentially represent spatially distributed time-series signals, the performance difference between these input formats can be explained by using this conclusion.
With magnitude-phase concatenation, the model can selectively focus more on the magnitude component while largely ignoring the less relevant phase information.
However, with real-imaginary concatenation, the magnitude and phase information become entangled in a joint probability distribution.
This joint probability distribution makes using the discriminative magnitude features more challenging for the model.
Further improvement was achieved when using only magnitude information as input, yielding the best performance with a linear probing error rate of 0.23\% (72\% relative improvement over magnitude-phase concatenation) and a fine-tuning error rate of 0.06\% (63\% relative improvement). 
These results demonstrate that the magnitude spectrum alone provides the most critical and discriminative information for event classification in DAS applications.

\section{Conclusion}
In conclusion, this work presents DAS-MAEv2, an enhanced framework that significantly advances representation learning for DAS through two key innovations: STFT transformation for explicit frequency feature extraction and cross-domain video pre-training.
By pioneering the first successful bridging of video sequences and waterfall plots in DAS applications, our approach establishes a new paradigm for learning highly transferable and capable representations that significantly outperform previous methods.
The model's state-of-the-art performance validates video data as effective pre-training material, which enables learning comprehensive spatial-temporal-frequency representations. 
These capabilities prove particularly valuable for industrial applications requiring sophisticated waterfall plot analysis.
Crucially, our framework enables data-efficient learning that aligns with IoT's requirements for flexible multimodal systems, reducing domain-specific data requirements through effective cross-modal transfer. 
This breakthrough not only advances DAS but establishes a transformative approach to representation learning with broad implications for industrial IoT applications.

\bibliography{report}

@article{radford2023robust,
  title={Robust speech recognition via large-scale weak supervision},
  author={Radford, Alec and Kim, Jong Wook and Xu, Tao and Brockman, Greg and McLeavey, Christine and Sutskever, Ilya},
  journal={International conference on machine learning},
  pages={28492--28518},
  year={2023},
  organization={PMLR}
}

@article{wang2023neural,
  title={Neural codec language models are zero-shot text to speech synthesizers},
  author={Wang, Chengyi and Chen, Sanyuan and Wu, Yu and Zhang, Ziqiang and Zhou, Long and Liu, Shujie and Chen, Zhuo and Liu, Yanqing and Wang, Huaming and Li, Jinyu and others},
  journal={arXiv preprint arXiv:2301.02111},
  year={2023}
}

@article{li2020fiber,
  title={Fiber distributed acoustic sensing using convolutional long short-term memory network: a field test on high-speed railway intrusion detection},
  author={Li, Zhongqi and Zhang, Jianwei and Wang, Maoning and Zhong, Yuzhong and Peng, Fei},
  journal={Optics express},
  volume={28},
  number={3},
  pages={2925--2938},
  year={2020},
  publisher={OSA}
}

@article{li2024exploiting,
  title={Exploiting CNN-BiLSTM Model for Distributed Acoustic Sensing Event Recognition},
  author={Li, Zhiheng},
  journal={2024 International Conference on Artificial Intelligence and Communication (ICAIC 2024)},
  pages={333--341},
  year={2024},
  organization={Atlantis Press}
}

@article{ma2022mi,
  title={MI-SI based distributed optical fiber sensor for no-blind zone location and pattern recognition},
  author={Ma, Yixiao and Song, Yuchen and Song, Qiuheng and Xiao, Qian and Jia, Bo},
  journal={Journal of Lightwave Technology},
  volume={40},
  number={9},
  pages={3022--3030},
  year={2022},
  publisher={IEEE}
}

@article{zhao2021markov,
  title={Markov transition fields and deep learning-based event-classification and vibration-frequency measurement for $\varphi$-OTDR},
  author={Zhao, Xiaoting and Sun, Hongbin and Lin, Bo and Zhao, Hongmin and Niu, Yingli and Zhong, Xiang and Wang, Yidan and Zhao, Yiming and Meng, Fanchao and Ding, Jinmin and others},
  journal={IEEE Sensors Journal},
  volume={22},
  number={4},
  pages={3348--3357},
  year={2021},
  publisher={IEEE}
}

@article{pan2022time,
  title={Time attention analysis method for vibration pattern recognition of distributed optic fiber sensor},
  author={Pan, Yining and Wen, Tingkun and Ye, Wei},
  journal={Optik},
  volume={251},
  pages={168127},
  year={2022},
  publisher={Elsevier}
}

@article{wu2019dynamic,
  title={A dynamic time sequence recognition and knowledge mining method based on the hidden Markov models (HMMs) for pipeline safety monitoring with $\Phi$-OTDR},
  author={Wu, Huijuan and Liu, Xiangrong and Xiao, Yao and Rao, Yunjiang},
  journal={Journal of Lightwave Technology},
  volume={37},
  number={19},
  pages={4991--5000},
  year={2019},
  publisher={OSA}
}

@article{tejedor2019contextual,
  title={A contextual GMM-HMM smart fiber optic surveillance system for pipeline integrity threat detection},
  author={Tejedor, Javier and Macias-Guarasa, Javier and Martins, Hugo F and Martin-Lopez, Sonia and Gonzalez-Herraez, Miguel},
  journal={Journal of Lightwave Technology},
  volume={37},
  number={18},
  pages={4514--4522},
  year={2019},
  publisher={OSA}
}

@article{yi2023intelligent,
  title={An intelligent crash recognition method based on 1DResNet-SVM with distributed vibration sensors},
  author={Yi, Jichao and Shang, Ying and Wang, Chen and Du, Yuankai and Yang, Jian and Sun, Maocheng and Huang, Sheng and Qu, Shuai and Zhao, Wenan and Zhao, Yanjie and others},
  journal={Optics Communications},
  volume={536},
  pages={129263},
  year={2023},
  publisher={Elsevier}
}

@article{yang2023using,
  title={Using phase-sensitive optical time domain reflectometers to develop an alignment-free end-to-end multitarget recognition model},
  author={Yang, Nachuan and Zhao, Yongjun and Wang, Fuqiang and Chen, Jinyang},
  journal={Electronics},
  volume={12},
  number={7},
  pages={1617},
  year={2023},
  publisher={MDPI}
}

@article{wu2021intelligent,
  title={Intelligent target recognition for distributed acoustic sensors by using both manual and deep features},
  author={Wu, Huijuan and Wang, Chaoqun and Liu, Xinyu and Gan, DengKe and Liu, Yimeng and Rao, Yunjiang and Olaribigbe, Abdulafeez Olawale},
  journal={Applied optics},
  volume={60},
  number={23},
  pages={6878--6887},
  year={2021},
  publisher={Optical Society of America}
}

@article{kayan2023intensity,
  title={Intensity and phase stacked analysis of a $\Phi$-OTDR system using deep transfer learning and recurrent neural networks},
  author={Kayan, Ceyhun Efe and Yuksel Aldogan, Kivilcim and Gumus, Abdurrahman},
  journal={Applied Optics},
  volume={62},
  number={7},
  pages={1753--1764},
  year={2023},
  publisher={Optica Publishing Group}
}

@article{rose2015internet,
  title={The internet of things: An overview},
  author={Rose, Karen and Eldridge, Scott and Chapin, Lyman and others},
  journal={The internet society (ISOC)},
  volume={80},
  number={15},
  pages={1--53},
  year={2015},
  publisher={Reston, VA}
}

@article{li2015internet,
  title={The internet of things: a survey},
  author={Li, Shancang and Xu, Li Da and Zhao, Shanshan},
  journal={Information systems frontiers},
  volume={17},
  number={2},
  pages={243--259},
  year={2015},
  publisher={Springer}
}

@inproceedings{vivit,
  title={Vivit: A video vision transformer},
  author={Arnab, Anurag and Dehghani, Mostafa and Heigold, Georg and Sun, Chen and Lu{\v{c}}i{\'c}, Mario and Schmid, Cordelia},
  booktitle={Proceedings of the IEEE/CVF international conference on computer vision},
  pages={6836--6846},
  year={2021}
}

@article{duan2025mae,
  title={DAS-MAE: A self-supervised pre-training framework for universal and high-performance representation learning of distributed fiber-optic acoustic sensing},
  author={Duan, Junyi and Chen, Jiageng and He, Zuyuan},
  journal={arXiv preprint arXiv:2506.04552},
  year={2025}
}

@article{ren2015theoretical,
  title={Theoretical and Experimental Analysis of $\Phi$-OTDR Based on Polarization Diversity Detection},
  author={Ren, Meiqi and Lu, Ping and Chen, Liang and Bao, Xiaoyi},
  journal={IEEE Photonics Technology Letters},
  volume={28},
  number={6},
  pages={697--700},
  year={2015},
  publisher={IEEE}
}

@article{chen2019108,
  title={108-km Distributed Acoustic Sensor With 220-p $epsilon/\surd $ Hz Strain Resolution and 5-m Spatial Resolution},
  author={Chen, Dian and Liu, Qingwen and He, Zuyuan},
  journal={Journal of Lightwave Technology},
  volume={37},
  number={18},
  pages={4462--4468},
  year={2019},
  publisher={IEEE}
}

@article{dong2016quantitative,
  title={Quantitative measurement of dynamic nanostrain based on a phase-sensitive optical time domain reflectometer},
  author={Dong, Yongkang and Chen, Xi and Liu, Erhu and Fu, Cheng and Zhang, Hongying and Lu, Zhiwei},
  journal={Applied optics},
  volume={55},
  number={28},
  pages={7810--7815},
  year={2016},
  publisher={Optical Society of America}
}

@ARTICLE{DAS,
  title={Contributed review: Distributed optical fibre dynamic strain sensing},
  author={Masoudi, Ali and Newson, Trevor P},
  journal={Review of scientific instruments},
  volume={87},
  number={1},
  year={2016},
  publisher={AIP Publishing}
}

@article{phai-otdr,
  title={Distributed vibration sensor based on coherent detection of phase-OTDR},
  author={Lu, Yuelan and Zhu, Tao and Chen, Liang and Bao, Xiaoyi},
  journal={Journal of lightwave Technology},
  volume={28},
  number={22},
  pages={3243--3249},
  year={2010},
  publisher={IEEE}
}

@ARTICLE{DAS2,
  author={He, Zuyuan and Liu, Qingwen},
  journal={Journal of Lightwave Technology}, 
  title={Optical Fiber Distributed Acoustic Sensors: A Review}, 
  year={2021},
  volume={39},
  number={12},
  pages={3671-3686},
  doi={10.1109/JLT.2021.3059771}}

@article{eq1,
  title={Convolutional neural network for earthquake detection and location},
  author={Perol, Thibaut and Gharbi, Micha{\"e}l and Denolle, Marine},
  journal={Science Advances},
  volume={4},
  number={2},
  pages={e1700578},
  year={2018},
  publisher={American Association for the Advancement of Science}
}

@article{eq2,
  title={Generalized seismic phase detection with deep learning},
  author={Ross, Zachary E and Meier, Men-Andrin and Hauksson, Egill and Heaton, Thomas H},
  journal={Bulletin of the Seismological Society of America},
  volume={108},
  number={5A},
  pages={2894--2901},
  year={2018},
  publisher={GeoScienceWorld}
}

@article{seismic,
  title={Seismic monitoring with distributed acoustic sensing from the near-surface to the deep oceans},
  author={Fern{\'a}ndez-Ruiz, Mar{\'\i}a R and Martins, Hugo F and Williams and Gonz{\'a}lez-Herr{\'a}ez, Miguel},
  journal={Journal of Lightwave Technology},
  volume={40},
  number={5},
  pages={1453--1463},
  year={2022},
  publisher={IEEE}
}

@article{pipeline1,
  title={Toward prevention of pipeline integrity threats using a smart fiber-optic surveillance system},
  author={Tejedor, Javier and Martins,  Willy and Gonz{\'a}lez-Herr{\'a}ez, Miguel},
  journal={Journal of Lightwave Technology},
  volume={34},
  number={19},
  pages={4445--4453},
  year={2016},
  publisher={IEEE}
}

@article{pipeline2,
  title={Inspection and monitoring systems subsea pipelines: A review paper},
  author={Ho, Michael and El-Borgi, Sami and Patil, Devendra and Song, Gangbing},
  journal={Structural Health Monitoring},
  volume={19},
  number={2},
  pages={606--645},
  year={2020},
  publisher={SAGE Publications Sage UK: London, England}
}

@article{railway1,
  title={A review of railway infrastructure monitoring using fiber optic sensors},
  author={Du, Cong and Dutta, Susom and Kurup, Pradeep and Yu, Tzuyang and Wang, Xingwei},
  journal={Sensors and Actuators A: Physical},
  volume={303},
  pages={111728},
  year={2020},
  publisher={Elsevier}
}

@article{railway2,
title = {A review of distributed acoustic sensing applications for railroad condition monitoring},
journal = {Mechanical Systems and Signal Processing},
volume = {208},
pages = {110983},
year = {2024},
issn = {0888-3270},
doi = {https://doi.org/10.1016/j.ymssp.2023.110983},
author = {Md Arifur Rahman and Hossein Taheri and Fadwa Dababneh and Sasan Sattarpanah Karganroudi and Seyyedabbas Arhamnamazi},
}

@article{MT,
  title={Mean teachers are better role models: Weight-averaged consistency targets improve semi-supervised deep learning results},
  author={Tarvainen, Antti and Valpola, Harri},
  journal={Advances in neural information processing systems},
  volume={30},
  year={2017}
}

@article{van2021self,
  title={A self-supervised deep learning approach for blind denoising and waveform coherence enhancement in distributed acoustic sensing data},
  author={Van den Ende, Martijn and Lior, Itzhak and Richard, C{\'e}dric},
  journal={IEEE Transactions on Neural Networks and Learning Systems},
  volume={34},
  number={7},
  pages={3371--3384},
  year={2021},
  publisher={IEEE}
}

@article{wu2021pattern,
  title={Pattern recognition in distributed fiber-optic acoustic sensor using an intensity and phase stacked convolutional neural network with data augmentation},
  author={Wu, Huan and Zhou, Bin and Zhu, Kun and Shang, Chao and Tam, Hwa-Yaw and Lu, Chao},
  journal={Optics express},
  volume={29},
  number={3},
  pages={3269--3283},
  year={2021},
  publisher={Optica Publishing Group}
}

@article{videomae,
  title={Videomae: Masked autoencoders are data-efficient learners for self-supervised video pre-training},
  author={Tong, Zhan and Song, Yibing and Wang, Jue and Wang, Limin},
  journal={Advances in neural information processing systems},
  volume={35},
  pages={10078--10093},
  year={2022}
}

@article{2017attention,
  title={Attention is all you need},
  author={Vaswani, A},
  journal={Advances in Neural Information Processing Systems},
  year={2017}
}

@article{wu1dcnn,
  title={One-dimensional CNN-based intelligent recognition of vibrations in pipeline monitoring with DAS},
  author={Wu, Huijuan and Chen, Jiping and Liu, Xiangrong and Xiao, Yao and Wang, Mengjiao and Zheng, Yi and Rao, Yunjiang},
  journal={Journal of Lightwave Technology},
  volume={37},
  number={17},
  pages={4359--4366},
  year={2019},
  publisher={IEEE}
}

@article{vit,
  title={An image is worth 16x16 words: Transformers for image recognition at scale},
  author={Dosovitskiy, Alexey and Beyer, Lucas and Kolesnikov, Alexander and Weissenborn, Dirk and Zhai, Xiaohua and Unterthiner, Thomas and Dehghani, Mostafa and Minderer, Matthias and Heigold, Georg and Gelly, Sylvain and others},
  journal={arXiv preprint arXiv:2010.11929},
  year={2020}
}

@article{MTssl,
  title={A deep learning model enabled multi-event recognition for distributed optical fiber sensing},
  author={Li, Yujiao and Cao, Xiaomin and Ni, Wenhao and Yu, Kuanglu},
  journal={Science China Information Sciences},
  volume={67},
  number={3},
  pages={132404},
  year={2024},
  publisher={Springer}
}

@article{dataset,
  title={An open dataset of $\varphi$-OTDR events with two classification models as baselines},
  author={Cao, Xiaomin and Su, Yunsheng and Jin, Zhiyan and Yu, Kuanglu},
  journal={Results in Optics},
  volume={10},
  pages={100372},
  year={2023},
  publisher={Elsevier}
}

@inproceedings{MAE,
  title={Masked autoencoders are scalable vision learners},
  author={He, Kaiming and Chen, Xinlei and Xie, Saining and Li, Yanghao and Doll{\'a}r, Piotr and Girshick, Ross},
  booktitle={Proceedings of the IEEE/CVF conference on computer vision and pattern recognition},
  pages={16000--16009},
  year={2022}
}

@article{krizhevsky2012imagenet,
  title={Imagenet classification with deep convolutional neural networks},
  author={Krizhevsky, Alex and Sutskever, Ilya and Hinton, Geoffrey E},
  journal={Advances in neural information processing systems},
  volume={25},
  year={2012}
}

@article{adamw,
  title={Decoupled weight decay regularization},
  author={Loshchilov, I},
  journal={arXiv preprint arXiv:1711.05101},
  year={2017}
}

@article{cos_lr,
  title={Sgdr: Stochastic gradient descent with warm restarts},
  author={Loshchilov, Ilya and Hutter, Frank},
  journal={arXiv preprint arXiv:1608.03983},
  year={2016}
}

@article{t-SNE,
  title={Visualizing data using t-SNE.},
  author={Van der Maaten, Laurens and Hinton, Geoffrey},
  journal={Journal of machine learning research},
  volume={9},
  number={11},
  year={2008}
}

@article{PCA,
  title={Principal components analysis (PCA)},
  author={Ma{\'c}kiewicz, Andrzej and Ratajczak, Waldemar},
  journal={Computers \& Geosciences},
  volume={19},
  number={3},
  pages={303--342},
  year={1993},
  publisher={Elsevier}
}

@article{wu2020novel,
  title={A novel DAS signal recognition method based on spatiotemporal information extraction with 1DCNNs-BiLSTM network},
  author={Wu, Huijuan and Yang, Mingru and Rao, Yunjiang},
  journal={IEEE Access},
  volume={8},
  pages={119448--119457},
  year={2020},
  publisher={IEEE}
}
\bibliographystyle{IEEEtran}
 
\vfill

\end{document}